\documentclass[pteplogo]{ptephy}

\preprintnumber{} 

\newtheorem{theorem}{Theorem}

\begin{document}

\title{QCD sum rules on the complex Borel plane}

\author{\name{Ken-Ji  Araki}{1,\ast}, \name{Keisuke Ohtani}{1}, \name{Philipp Gubler}{2}, \name{Makoto Oka}{1}   }


\address{\affil{1}{Department of Physics, H-27, Tokyo Institute of Technology,
Meguro, Tokyo 152-8551, Japan
}
\affil{2}{RIKEN Nishina Center, 2-1 Hirosawa, Wako 351-0106, Japan}
\email{k.araki@th.phys.titech.ac.jp}}

\begin{abstract}%
Borel transformed QCD sum rules conventionally use a real valued parameter 
(the Borel mass) for specifying the exponential weight over which hadronic 
spectral functions are averaged. 
In this paper, it is shown that the Borel mass can be generalized to have 
complex values and that new classes of sum rules can be derived 
from the resulting averages over the spectral functions. 
The real and imaginary parts of 
these novel sum rules turn out to have damped oscillating kernels and 
potentially contain a larger amount of information on the hadronic 
spectrum than the real valued QCD sum rules. 
As a first practical test, 
we have formulated the complex Borel sum rules for the $\phi$ 
meson channel and have analyzed them using the maximum entropy 
method, by which we can extract the most probable spectral function from 
the sum rules without strong assumptions on its functional form. 
As a result, it is demonstrated that, compared to earlier studies, the 
complex valued sum rules allow us to extract 
the spectral function with a significantly improved resolution and 
thus to study more detailed structures of the hadronic spectrum than 
previously possible.  

\end{abstract}

\subjectindex{D32, B67 }

\maketitle

\section{Introduction \label{sec:introduction}}

The spectral function of hadrons is one of the main targets in studies of low energy QCD. 
At low energy, non-perturbative approaches are inevitable as 
the coupling constant is large and 
the QCD vacuum has non-trivial quark and gluon condensates.
In order to take into account effects of these vacuum condensates, QCD sum rules 
\cite{Shifman1979,Shifman1979-2,Reinders1985} have been extensively used to 
explore hadron spectra. 

QCD sum rules utilize the operator product expansion (OPE) for evaluating correlation functions, which is valid in the deep Euclidean four-momentum region. 
A dispersion relation based on analyticity of the correlation function on the other hand 
yields a relation between an integral over the spectral function and the vacuum condensates. 
Inverting the integral relation and extracting the spectral function thus is the central issue of QCD sum rule analyses. 
In conventional approaches, the spectral function is most often parametrized using a 
``pole + continuum'' functional form, whose parameters are determined to satisfy the sum rule. 
This technique is, however, not always applicable 
because in reality the spectral functions are not restricted to a particular shape. 

Recently, a new method was proposed that directly provides the spectral function without assuming a functional shape \cite{Gubler2010}. 
It utilizes the maximum entropy method (MEM), which generally 
helps to determine the most probable spectral function from 
an integral relation \cite{Asakawa2001}. 
Thus, the obtained spectral function is chosen from infinitely many functional forms, while the conventional approach 
only gives the best fitted ``pole + continuum'' type function. 
So far, this novel method has been applied to the $\rho$-meson \cite{Gubler2010} and nucleon \cite{Ohtani2011,Ohtani2013} channels 
in vacuum and to charmonium \cite{Gubler2011} and bottomonium \cite{Suzuki2013} channels at finite temperature. 
It has, however, not yet 
shown its full strength, giving only the ground state peak structure, while usually neither reproducing excitation nor continuum spectra. 
We believe that this is not a consequence of the limitation of MEM, but rather due to the limited information provided by 
the conventional QCD sum rules. 

In this paper, we propose to extend the QCD sum rules to the complex plane of the squared-momentum, $z=q^2$, 
by which we are able to extract more information on the spectral function.\footnote{%
Extension of QCD to the complex $q^2$ plane has been proposed earlier by Ioffe and Zyablyuk \cite{Ioffe2001}.} 
As the sum rules are based on the analytic continuation of the correlation function on the $q^2$ plane, 
they can be naturally generalized to the complex plane. 
As a result, it is found that after using 
the Borel transform to enhance the convergence of the OPE, 
the Borel sum rule is valid also for the complex Borel parameter. 

Applying the MEM to the newly constructed sum rules, we study the spectral function of the vector meson composed of the strange quark ($s \bar s$), i.e. 
the spectral function in the $\phi$ meson channel. 
Our results show that the new sum rule improves the reproducibility of the 
spectral function compared to the conventional Borel sum rules,  
in particular in the large momentum region. 

The paper is organized as follows. 
In Sec.\ref{s:comp}, we explain the central idea of our novel complex Borel plane sum rules, demonstrate in detail 
how the sum rules are constructed and discuss their properties. 
After all, it will be shown that our formulation can be considered to be just a 
simple generalization of conventional Borel sum rules to complex Borel mass values. 
Next, we briefly introduce MEM and define the likelihood function for the complex Borel plane sum rules to apply MEM to them in Sec. \ref{s:max}.
In  Sec. \ref{s:ana} the results of the analyses are outlined. 
Here, we will 
not only show the results of the complex Borel plane sum rules but also the ones of the original Borel sum rules 
for comparison. 
Summary and conclusions are given in Sec. \ref{s:sum}. 

\section{Complex Borel sum rules}
\label{s:comp} 
In this section, we formulate the complex Borel sum rules (CBSR). 
The general procedure is the same as that used for deriving the conventional QCD sum rules with real variables (RBSR).

\subsection{Dispersion relation on the complex plane}
The basic idea of the CBSR is to consider the squared four-momentum $q^2$ as a complex parameter 
when deriving the sum rules. 
The validity of this generalization
is guaranteed by the dispersion relation. 

\begin{figure}[!tbp]
 \begin{center}
  \includegraphics[scale=0.4]{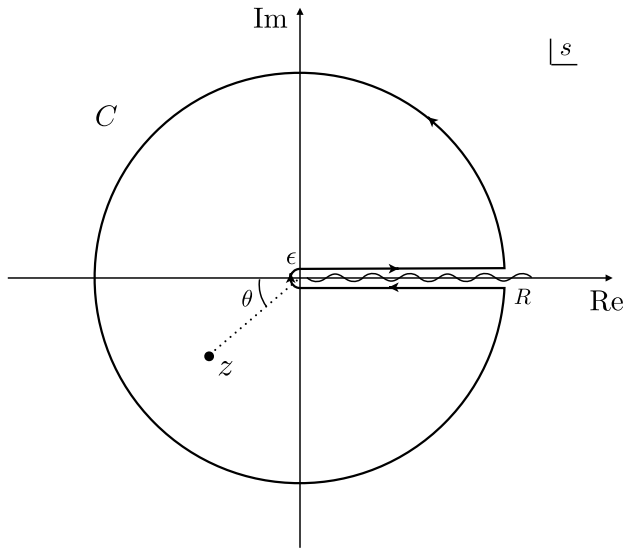}
 \end{center}
 \caption{The contour integral $C$ on the complex plane of the variable $s$, used 
in Eq.(\ref{eq:sumrulestart}). For the actual calculations, the radius of the outer circle of 
$C$ is taken to infinity. 
The wavy line denotes the non-analytic cut (or poles) of $\Pi(s)$ on the 
positive side of the real axis. }
 \label{fig:contour2}
\end{figure}

The elementary ingredients of the derivation, Cauchy's residue theorem and the analyticity of the correlator, 
do not restrict $q^2$ to real numbers but rather allow it to 
have any value on the 
complex plane except the region around the positive real axis (see Fig. \ref{fig:contour2}). 
In other words, for the correlator $\Pi(q^2)$, Cauchy's residue theorem and analyticity guarantee
\begin{equation}
\Pi(z) = \frac{1}{2\pi i}\oint _{C} \frac{z^n \Pi(s)}{s^n(s-z)}ds,
\label{eq:sumrulestart}
\end{equation}
for any complex $z=q^2$. 
Here, $n$ is a positive integer, chosen to be large enough for the integral to converge. 
The detailed definition of the $\Pi(q^2)$ depends on the specific channel. In the case of the vector channel, which will be studied 
in this paper, $\Pi(q^2)$ can be defined as shown in Eq.(21). 
Following the same steps used when deriving the conventional sum rules, 
we obtain the following dispersion relation (see Appendix A for details): 
\begin{equation}
\Pi(z) = \int_{0}^{\infty} \frac{\rho(s)}{s-z}ds + \hbox{(polynomial in $z$)},
\label{eq:cdis}
\end{equation}
where $\rho(s)= \frac{1}{\pi} {\rm Im}\, \Pi (s+i\epsilon)$ (for a real $s$ and an infinitesimal $\epsilon$ ($>0$)) is the spectral function.
Although this looks just like the conventional diversion relation, it potentially contains novel kinds of sum rules for the spectral function, 
extracted from both its real and imaginary parts. 
Our strategy is employing Eq.(\ref{eq:cdis}) as 
a starting point for deriving the actual sum rules. 

\subsection{Analytic continuation of the OPE}
\label{subsec:3.2} 
In QCD sum rules, it is common to replace the left hand side of Eq.(\ref{eq:cdis}) by its respective OPE, which 
is valid in the deep Euclidean region, i.e. large real $(-q^2)$. 
Extending this to the complex plane, we consider the OPE for the complex variable $z=q^2$. 
In practice, the OPE must be truncated at a certain operator dimension and 
one can only hope that it converges at sufficiently large $(-q^2)$. 
In the region where the OPE is convergent, the left hand side of Eq.(\ref{eq:cdis}) can be extended to the complex argument $z$ as $\Pi^{\mathrm{OPE}}(z)$, 
which is the 
analytically continued function of $\Pi^{\mathrm{OPE}}(q^2)$. 
It is important to note that $\Pi^{\mathrm{OPE}}(q^2)$ depends on $q^2$ only through the Wilson coefficients and therefore 
has the same vacuum expectation values of the local operators. 
After all, we obtain the complex sum rules as, 
\begin{equation}
\Pi^{\mathrm{OPE}}(z) = \int_{0}^{\infty} \frac{\rho(s)}{s-z}ds + \hbox{(polynomial in $z$)}.
\label{eq:cdis2}
\end{equation}

\subsection{Borel transformation}
\label{effectivity}
The unknown polynomial in Eq.(\ref{eq:cdis2}) can be removed by the Borel transform,
defined by
\begin{equation}
\begin{split}
\hat{B}_{[X]}=\lim_{\stackrel {X,n \rightarrow \infty}{X/n=M^2}  }\frac{X^n}{(n-1)!}  \Bigl(   - \frac{\partial}{\partial X}   \Bigr)^n, 
\end{split}
\label{eq:oldfash}
\end{equation}
where $X$ is a real variable. 

By substituting $z=|z|\mathrm{e}^{i(\theta-\pi)}$, where $\theta$ is defined as shown in Fig. \ref{fig:contour2},
Eq.(\ref{eq:cdis2}) can be considered as a relation depending on $|z|$ and $\theta$.
As polynomials in z are linear combinations of $|z|^k \mathrm{e}^{i k (\theta - \pi)}$, differentiating Eq.(\ref{eq:cdis2}) infinite times by $|z|$ 
eliminates them. It is hence understood that $\hat{B}_{[|z|]}$ is suitable for our present purposes. 
On the right hand side of Eq.(\ref{eq:cdis2}), the integral term is transformed as
\begin{equation}
\begin{split}
\hat{B}_{[|z|]} \int_{0}^{\infty} \frac{\rho(s)}{s-z}ds&=\lim_{\stackrel {|z|,n \rightarrow \infty}{|z|/n=M^2}  }\frac{|z|^n}{(n-1)!}\Bigl(   - \frac{\partial}{\partial |z|}   \Bigr)^n\int_{0}^{\infty} \frac{\rho(s)}{s-z}ds\\
&=   \lim_{\stackrel {|z|,n \rightarrow \infty}{|z|/n=M^2}  }\frac{|z|^n}{(n-1)!}   \int_{0}^{\infty}\frac{n! \mathrm{e}^{in\theta}}{(s+|z|\mathrm{e}^{i\theta})^{n+1}}\rho(s)ds\\
&= \lim_{\stackrel {|z|,n \rightarrow \infty}{|z|/n=M^2}  }  \frac{n}{|z|\mathrm{e}^{i\theta} }  \int_{0}^{\infty}   \Bigl(  \frac{  |z|\mathrm{e}^{i\theta}   }{s+|z|\mathrm{e}^{i\theta}} \Bigr)^{n+1}\rho(s)ds\\
&=\lim_{n \to \infty} \frac{1}{M^2\mathrm{e}^{i\theta} }\int_{0}^{\infty}  \Bigl(  1  +  \frac{1}{n}\frac{s}{M^2\mathrm{e}^{i\theta}}   \Bigr)^{-(n+1)}\rho(s)ds.
\end{split}
\label{eq:keisan}
\end{equation}
The integral kernel is transformed into an exponential function 
\textit{if} the limit $n \to \infty$ can be interchanged with the integral over $s$. 
At first sight, this seems to be a trivially allowed manipulation, 
but a careful inspection, in fact, shows that it is not necessarily correct. 
Relying on 
a theorem (which is similar to Lebesgue's ``dominated convergence theorem"), 
it is possible to 
show that the two operations indeed can be interchanged 
for $\cos{\theta}>0$. 
An explicit proof of this statement is given in Appendix B. 
On the other hand, for $\cos\theta<0$, one sees that the $n\to\infty$ limit and the integral over s are not interchangeable. 
This is so because if one could interchange the limit with the integral, 
the ensuing integrand would diverge  
exponentially at large $s\to\infty$, which is apparently inconsistent with the left hand side, 
which remains finite for $\cos\theta<0$, as we will see in the discussion given below. 
We thus conclude that only for $\cos{\theta}>0$, the Borel transform leads to the following 
result: 
\begin{equation}
 \ \ \ \ \ \ \ \hat{B}_{[|z|]} \int_{0}^{\infty} \frac{\rho(s)}{s-z}ds=\frac{1}{ M^2\mathrm{e}^{i\theta} }\int^{\infty}_0  
 \mathrm{e}^{-s/( M^2 \mathrm{e}^{i\theta})} \rho (s) ds.
\end{equation}
Note that this region includes $\cos\theta=1$ ($\theta=0$), which gives the RBSR.

Next, let us discuss the Borel transformation of the left hand side of Eq.(\ref{eq:cdis2}). 
As above, we substitute $z=|z|\mathrm{e}^{i(\theta-\pi)}$ 
into the given analytically continued OPE expression and then apply $\hat{B}_{[|z|]}$. 
Doing this, we obtain 
\begin{equation}
\begin{split}
\hat{B}_{[|z|]}\,z^k&=0,\\
\ \hat{B}_{[|z|]} \,\Bigl( \frac{1}{z} \Bigr)^k &=\frac{(-1)^k}{(k-1)!}\Bigl(\frac{1}{M^2\mathrm{e}^{i\theta}} \Bigr)^k,\\
\hat{B}_{[|z|]} \,z^k \ln{\Bigl(-\frac{z}{{\mu}^2} \Bigr)} &= -k!(M^2\mathrm{e}^{i\theta})^k,\\
\hat{B}_{[|z|]} \,\Bigl(\frac{1}{s-z}\Bigr)^k &=\frac{1}{(k-1)!}\frac{1}{(M^2\mathrm{e}^{i\theta})^k}  \mathrm{e}^{-s/( M^2 \mathrm{e}^{i\theta})}, 
\end{split}
\end{equation}
for which detailed 
derivations are given in Appendix C. 
Here let us compare these results with the following pre-existing formulae for the corresponding real functions: 
\begin{equation}
\begin{split}
\hat{B}_{[-q^2]}\,(q^2)^k&=0,\\
\hat{B}_{[-q^2]} \,\Bigl( \frac{1}{q^2} \Bigr)^k &=\frac{(-1)^k}{(k-1)!}\Bigl(\frac{1}{M^2}\Bigr)^k,\\
\hat{B}_{[-q^2]} \,(q^2)^k \ln{\Bigl(-\frac{q^2}{{\mu}^2}\Bigr)} &= -k!(M^2)^k, \\
\hat{B}_{[-q^2]} \,\Bigl(\frac{1}{s-q^2}\Bigr)^k &=\frac{1}{(k-1)!}\frac{1}{(M^2)^k}  \mathrm{e}^{-s/( M^2 )}. 
\end{split}
\label{eq:fr}
\end{equation}
These all suggest that the Borel transformation of the 
analytically continued of OPE equals that of the original OPE 
with a complex valued Borel mass. 
We can thus set 
\begin{equation}
 \hat{B}_{[|z|]}\Pi(z)=G^{\mathrm{OPE} }(M^2\mathrm{e}^{i\theta}),  
\end{equation}
where $G^{\mathrm{OPE}}(M^2)$ is defined as $G^{\mathrm{OPE}}(M^2) \equiv \hat{B}_{[-q^2]}\Pi^{\mathrm{OPE}}(q^2)$. 

Finally, we obtain 
\begin{equation}
G^{\mathrm{OPE}}( \mathcal{M}^2)=\frac{1}{ \mathcal{M}^2}\int^{\infty}_0  \mathrm{e}^{-s/ \mathcal{M}^2}\rho (s) ds    \ \ \ \ \ \ \ ( \mathrm{Re}[ \mathcal{M}^2 ] >0)
\label{eq:cp}
\end{equation}
where $\mathcal{M}^2 \equiv M^2 \mathrm{e}^{i\theta}$. 
This form of the complex plane QCD sum rule is nothing but the well known real valued Borel sum rule, 
in which the real Borel mass is replaced by its complex  analogue, $ \mathcal{M}^2$. 
Therefore, the complex Borel plane sum rules is found to be a simple generalization of the
ordinary Borel sum rules to the complex Borel mass plane and of course includes the latter at $\theta =0$. 

\subsection{Properties of the CBSR}
Although the CBSR of Eq.(\ref{eq:cp}) 
looks similar to its real counterpart, its content  is quite different. 
Since Eq.(\ref{eq:cp}) is complex valued, it simultaneously gives two sum rules which can be obtained from its real and imaginary part. 
Specifically, we have 
\begin{eqnarray}
\begin{cases}
\mathrm{Re}[G^{\mathrm{OPE}}(\mathcal{M}^2)]& =\scalebox{0.9}{$ \displaystyle\int^{\infty}_0 $}K^{\mathrm{R}}(\mathcal{M}^2;s)  \rho (s) ds, \\
\mathrm{Im}[G^{\mathrm{OPE}}(\mathcal{M}^2)] &=  \scalebox{0.9}{$\displaystyle\int^{\infty}_0$} K^{\mathrm{I}}(\mathcal{M}^2;s)  \rho (s) ds, 
\end{cases}
\label{eq:cp2}
\end{eqnarray}
where $ K^{\mathrm{R}}(\mathcal{M}^2;s)$ and $K^{\mathrm{I}}(\mathcal{M}^2 ;s)$ are defined as
\begin{equation}
\begin{split}
K^{\mathrm{R}}(\mathcal{M}^2;s) & \equiv \mathrm{Re}\Bigl[ \, \frac{1}{\mathcal{M}^2} \, \mathrm{e}^{-s/\mathcal{M}^2} \, \Bigr]\\
& =\frac{1}{M^2}  \mathrm{e}^{-( \cos{\theta}/M^2)s }  \cos{\bigl[ \, (\sin{\theta} /M^2)s - \theta \, \bigr]}, 
\label{eq:reker}
\end{split}
\end{equation}

\begin{equation}
\begin{split}
K^{\mathrm{I}}(\mathcal{M}^2;s) &  \equiv \mathrm{Im}\Bigl[  \, \frac{1}{\mathcal{M}^2} \, \mathrm{e}^{-s/\mathcal{M}^2} \, \Bigr]\\
&=\frac{1}{M^2}  \mathrm{e}^{-( \cos{\theta}/M^2)s }  \sin{\bigl[ \, (\sin{\theta} /M^2)s - \theta \, \bigr]}.
\label{eq:imker}
\end{split}
\end{equation}
\begin{figure}[!t]
 \begin{center}
  \includegraphics[scale=1.1]{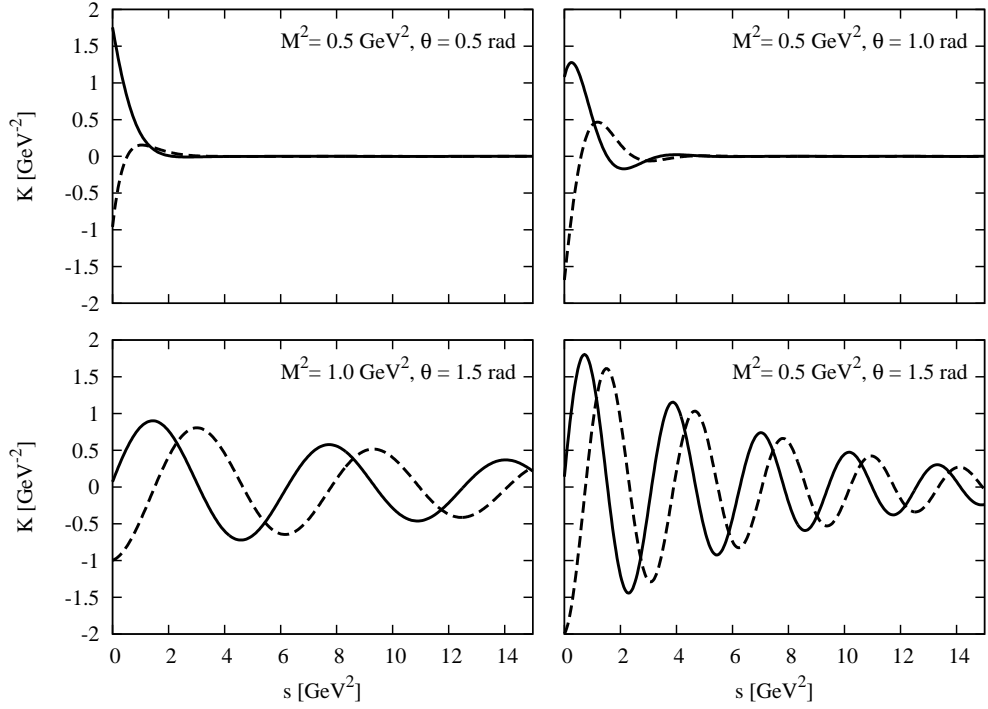}
 \end{center}
 \caption{The kernels $K^{\mathrm{R}}(\mathcal{M}^2;s)$ (solid lines) and $K^{\mathrm{I}}(\mathcal{M}^2;s)$ (dashed lines), 
 shown as a function of $s$, for various values of $\theta$ and $M^2$. }
 \label{fig:kernels}
\end{figure}
Both $K^{\mathrm{R}}$ and $K^{\mathrm{I}}$ are damped oscillating functions of $s$, as shown in Fig. \ref{fig:kernels} 
for several combinations of $M^2$ and $\theta$. As can be observed in these plots, the 
oscillations have various frequencies, phases and damping factors depending on the values of $M^2$ and $\theta$. 
We can hence expect that, compared with the RBSR, 
the sum rules with these kernels have the potential to resolve finer structures of the spectral function. 

As a further point, let us note here 
that the sum rules with complex Borel masses and their complex conjugates are not independent. 
Looking at the explicit form of the kernels, it is clear that the right hand sides of the sum rules with complex conjugated Borel masses satisfy 
\begin{equation}
\begin{cases}
\scalebox{0.9}{$ \displaystyle\int^{\infty}_0$} K^{\mathrm{R}}(  \overline{ \mathcal{M}^2};s)  \rho (s) ds &=  \scalebox{0.9}{$\displaystyle\int ^{\infty}_0 $} K^{\mathrm{R}}(  \mathcal{M}^2;s)  \rho (s) ds, \\
\scalebox{0.9}{$  \displaystyle\int^{\infty}_0$} K^{\mathrm{I}}(\overline{ \mathcal{M}^2 };s)  \rho (s) ds &=  - \scalebox{0.9}{$\displaystyle\int^{\infty}_0 $}K^{\mathrm{I}}( \mathcal{M}^2 ;s)  \rho (s) ds. 
\end{cases}
\end{equation}
In turn, the left hand side satisfies, according to the Schwarz reflection principle,
\begin{equation}
G^{\mathrm{OPE}}(\overline{\mathcal{M}^2})=\overline{G^{\mathrm{OPE}}(\mathcal{M}^2)}. 
\label{eq:conjugate2}
\end{equation}
Therefore the complex Borel sum rules are in essence identical to their 
complex conjugated counterparts. 

\subsection{\label{domain} Effective domain in complex Borel space}
It is important to specify the region on the complex Borel plane, in which the CBSR of 
Eq.(\ref{eq:cp}) (or Eq.(\ref{eq:cp2})) 
can be effectively used for the analysis of the spectral function. 
Firstly, the CBSR does not work for $\mathrm{Re}[\mathcal{M}^2] \leq 0$, 
as we have explained in section \ref{effectivity}. 
Secondly, the region with $\mathrm{Im}[\mathcal{M}^2] < 0$ 
gives sum rules with the same content as those with positive imaginary part of $\mathcal{M}^2$. 
Thirdly, we have to make sure to exclude the region, where the OPE 
might not be a valid approximation. 
To this end, we use the condition employed 
in standard QCD sum rule analyses, 
namely, we demand that the highest order OPE 
term is smaller than a critical ratio $r_c$ of the whole OPE expression. We thus impose 
\begin{equation}
\frac{|d^{\rm{max}}(\mathcal{M}^2 )|}{|G^{\rm{OPE}}(\mathcal{M}^2)|} < r_c, 
\label{eq:criterion}
\end{equation}
which is used in the RBSR analyses with a typical value of $r_c=0.1$. 
For the present work, we will employ the same value. 
The only difference to the RBSR is that we here take ratios 
of moduli of complex valued functions instead of real ones. 
Eq.(\ref{eq:criterion}) produces a closed curve in the complex Borel mass plane 
in whose inner region the sum rules can not be used. 

With all these restrictions, the effective domain for the sum rule lies in the first quadrant of the $\mathcal{M}^2$ imaginary plane 
with an excluded small $|\mathcal{M}^2|$ region, as shown schematically in Fig. 3.

\section{The maximum entropy method}
\label{s:max} 
Advantages of the CBSR
can be most efficiently exploited 
with the help of the maximum entropy method (MEM).
This method enables us to determine the 
spectral function from the sum rules
without assuming a specific functional form such as 
the popular ``pole + continuum" ansatz. 
Therefore, the more detailed the available information from the OPE is, 
the more realistic and accurate the spectral function obtained by MEM will be. 
Conversely, without enough physical information, the extracted spectral function 
will depend strongly on the ``default model", which is an input 
of the MEM analysis, as will be explained below.
In this sense, the CBSR is very useful for the MEM analysis 
as it provides more physical information thanks to the rich structure of its kernels. 
In the next few paragraphs, we shall briefly explain the essence of MEM and 
point out some issues specific to the analysis presented in this paper. 

MEM will lead us to the spectral function $\rho$ that maximizes
\begin{equation}
Q[\rho ]= \alpha S[\rho ]-L[\rho ],      
\end{equation}
where
$S[\rho]$ stands for the Shannon-Jaynes entropy, defined by
\begin{equation}
 S[\rho ] = \int^{\infty}_{0} d \omega[  \rho(\omega)-m(\omega) -\rho(\omega) \log \frac{\rho(\omega)}{m(\omega)}   ].
\label{eq:Shannon}
\end{equation}
Here, $m(\omega)$ is some positive definite function called default model.
(Note that we change the variable from 
$s$ to $\omega=\sqrt{s}$ in the following.)
$S[\rho]$ ensures the positive definiteness of the spectral function and 
takes the maximum value at $\rho(\omega) = m(\omega)$.

$L[\rho]$ is the likelihood function, which 
incorporates physical information provided by the sum rule.
In the case of the real valued Borel sum rules, it is expressed as 
 \begin{equation}
 L[\rho]  =\frac{1}{2 N }\sum_i \frac{1}{\sigma^2_i}
  \left|G^{OPE}(M^2_i) -  \int^{\infty}_{0} d\omega \,  \frac{2\omega}{M_i^2}
 \mathrm{e}^{-\omega ^2/M_i^2}\rho(\omega)      \right|^2 , 
 \label{eq:likelihood}
\end{equation}
where the subscript $i$ ($=1, \ldots N$) specifies the discretized Borel mass and $\sigma _i$ is 
the error of the OPE data $G^{\mathrm{OPE}}(M^2_i)$.
The error is determined from the uncertainty of the vacuum condensates. 
The factor $2\omega$, 
appearing in the integrand of Eq.(\ref{eq:likelihood}) is 
a result of the variable change from $s$ to $\omega$. 

The parameter $\alpha$ is a positive real number
on which the spectral function maximizing $Q[\rho]$ depends. 
The range of this parameter is determined by MEM and we take 
a weighted average of 
the obtained spectral functions over $\alpha$ to get the final solution. 
For the details of this procedure, we refer the interested reader to \cite{Gubler2010}. 

In the application of MEM to the complex Borel plane sum rules,
some modification in  
the likelihood function is necessary. 
As one complex Borel mass gives two sum rules, the real and imaginary parts,
the likelihood function can be expressed as the sum of them as
\begin{equation}
\begin{split}
 L[\rho]  =\frac{1}{2 N    } \sum_{(i,j)} & 
 \left[   \frac{1}{\sigma _{ij} ^{R\, 2}} \left|   \mathrm{Re}[ G^{\mathrm{OPE}} (\mathcal{M}^2_{ij}  )]  -\int^{\infty}_{0} d\omega\, 2\omega K^{R}(\mathcal{M}^2_{ij};  \omega ^2  )\rho(\omega ) \right|^2 \right.    \\
& \left. + \frac{1}{\sigma _{ij} ^{I\,  2}} 
\left|   \mathrm{Im}[ G^{\mathrm{OPE}}( \mathcal{M}^2_{ij})]    -\int^{\infty}_{0}d\omega\, 2\omega  K^{I}( \mathcal{M}_{ij} ^2 ;\omega^2)\rho(\omega) \right|^2   \right].
\end{split}
\end{equation}
\begin{figure}[!tbp]
 \begin{center}
  \includegraphics[scale=0.45]{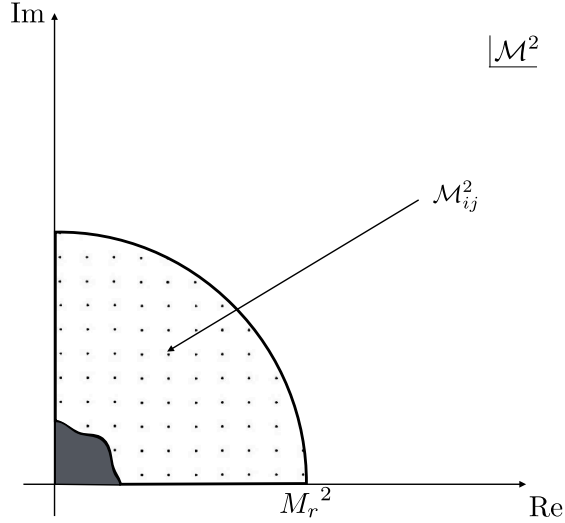}
 \end{center}
 \caption{A schematic illustration of the data points to be used in the MEM analysis and 
their respective distribution on the complex plane of the squared Borel mass $\mathcal{M}^2$.}
 \label{fig:floor}
\end{figure}
The subscripts $ij$ here specify 
discretized the complex Borel mass $\mathcal{M}^2_{ij}$
in the (2-dimensional) complex plane, and
$N$ is the total number of the chosen Borel masses.
The variances $\sigma^{\mathrm{R}} _{ij}$ and  $\sigma^{\mathrm{I}} _{ij}$ are
the ambiguities of the real and imaginary parts of $G^{\mathrm{OPE}}(\mathcal{M}^2_{ij})$. 

The discretized Borel masses are chosen in the domain shown in Fig.~\ref{fig:floor},
according to the previous discussion.
The lower boundary of $|\mathcal{M}^2|$ is determined 
to satisfy Eq.(\ref{eq:criterion}) for each $\theta$ independently. 
For the upper boundary, $|M_r^2|$, we do not have a definite restriction and
in principle may choose it freely.
We will show dependences on different choices of $M_r^2$ in the specific example 
given in the next section.

\section{Analyses of OPE data}
\label{s:ana} 
In this section, the CBSR is applied to the analysis of the spectral function for the $\phi$ meson channel 
as a first test of the validity of our method. 
We compare the results with those of the RBSR. 

\subsection{The CBSR for the $\phi$ meson}

We consider the sum rule for the vector meson composed of $s$ and $\bar s$, where $s$ is the strange quark. 
The interpolating field operator is $J^{\mu}(x)=\bar s(x) \gamma^{\mu} s(x)$, which is
supposed to create the $\phi$ (1020) meson from the vacuum.

The correlation function
\begin{equation}
\begin{split}
\Pi^{\mu\nu}(q^2)  &=i \int d^4 x \, e^{iq\cdot x} \langle 0| T(J^{\mu}(x) J^{\nu}(0)|0\rangle \\
&= (q^{\mu}q^{\nu} - g^{\mu\nu} q^2) \Pi(q^2)
\end{split}
\end{equation}
describes the spectrum of the $\phi$ meson and its excited states.

The OPE of the function $\Pi(q^2)$ has been obtained \cite{Shifman1979-2} as follows: 
\begin{equation}
\begin{split}
\Pi^{\mathrm{OPE}}(q^2)=&-\frac{1}{4 \pi ^2}(1+\frac{\alpha _s}{\pi})\ln (-\frac{q^2}{\mu ^2}) 
+\frac{3 m_s ^2}{2 \pi ^2} \frac{1}{q^2} + 2 m_s   \langle \bar{s } s \rangle \frac{1}{q^4}  \\
 &  +\frac{\langle\frac{ \alpha_s}{\pi}G ^2\rangle}{12} \frac{1}{q^4}   + \frac{224\pi \alpha_s}{81}     \kappa    \langle    \bar s s   \rangle ^2 \frac{1}{q^6}.
\end{split}
\label{eq:piOPE}
\end{equation}
Note that in writing down the above result, we have assumed the vacuum saturation approximation 
for the dimension 6 four-quark condensate term and have parametrized the 
possible violation of this approximation by the parameter $\kappa$. 
Replacing $q^2$ with $z=|z|\mathrm{e}^{i(\pi-\theta)}$ in Eq.(\ref{eq:piOPE}) and performing the 
Borel transformation, we can easily derive the OPE for the CBSR: 
\begin{equation}
\begin{split}
G^{\mathrm{OPE}}(\mathcal{M}^2)=&\frac{1}{4 \pi ^2}(1+\frac{\alpha _s}{\pi})     -\frac{3 m_s ^2}{2 \pi ^2} \frac{1}{\mathcal{M}^2} + 2 m_s   \langle \bar{s } s \rangle \frac{1}{\mathcal{M}^4}  \\
 &  +\frac{\langle\frac{ \alpha_s}{\pi}G ^2\rangle}{12} \frac{1}{\mathcal{M}^4}   - \frac{112\pi \alpha_s}{81}     \kappa   \langle    \bar s s   \rangle ^2 \frac{1}{\mathcal{M}^6},
\end{split}
\end{equation}
where $\mathcal{M}^2=M^2\mathrm{e}^{i\theta}$. As we mentioned in Sec.\ref{s:comp}, this form is equal to that of the real valued Borel sum rule, 
the Borel mass being simply replaced by the complex Borel mass. 
When we use the polar form for $\mathcal{M}^2$, the 
respective real and imaginary parts can be 
explicitly given as follows: 
\begin{equation}
\begin{cases}
\ \  \scalebox{0.98}{$\displaystyle\frac{1}{4 \pi ^2}(1+\frac{\alpha _s}{\pi})     -\frac{3 m_s ^2}{2 \pi ^2} \frac{\cos{ \theta}}{M^2} + 2 m_s   \langle \bar{s } s \rangle \frac{\cos{2\theta} }{M^4}
+\frac{\langle\frac{ \alpha_s}{\pi}G ^2\rangle}{12} \frac{\cos{2\theta}}{M^4}   - \frac{112\pi \alpha_s}{81}    \kappa  \langle    \bar s s   \rangle ^2 \frac{\cos{3\theta}}{M^6} $} \\
\ \ \ \ \ \ \ \ \ \ \ \ \ \ \ \ \ \ \ \ \ \ \ \ \ \ \ \ =\displaystyle\frac{1}{M^2} \scalebox{0.9}{$\displaystyle\int_{0}^{\infty}$} \mathrm{e}^{-( \cos{\theta}/M^2)s }  \cos{\bigl[ \, (\sin{\theta} /M^2)s - \theta \, \bigr]} \rho(s)ds, \\
\\
\ \  \scalebox{0.98}{$\displaystyle\frac{3 m_s ^2}{2 \pi ^2} \frac{\sin{ \theta}}{M^2} -2 m_s   \langle \bar{s } s \rangle \frac{\sin{2\theta} }{M^4} 
  -\frac{\langle\frac{ \alpha_s}{\pi}G ^2\rangle}{12} \frac{\sin{2\theta}}{M^4}   +\frac{112\pi \alpha_s}{81}      \kappa       \langle    \bar s s   \rangle ^2 \frac{\sin{3\theta}}{M^6} $}\\
\ \ \ \ \ \ \ \ \ \ \ \ \ \ \ \ \ \ \ \ \ \ \ \ \ \ \ \ =\displaystyle\frac{1}{M^2}  \scalebox{0.9}{$ \displaystyle\int_{0}^{\infty}$} \mathrm{e}^{-( \cos{\theta}/M^2)s }  \sin{\bigl[ \, (\sin{\theta} /M^2)s - \theta \, \bigr]} \rho(s)ds.
\end{cases}
\label{eq:realimag}
\end{equation}
The values and uncertainties of the quark mass, strong coupling constant and and vacuum condensates appearing in the OPE 
used in our analysis are given 
in Table \ref{table:condensate}.

\begin{figure*}[!h]
 \begin{center}
 \includegraphics[width=10cm]{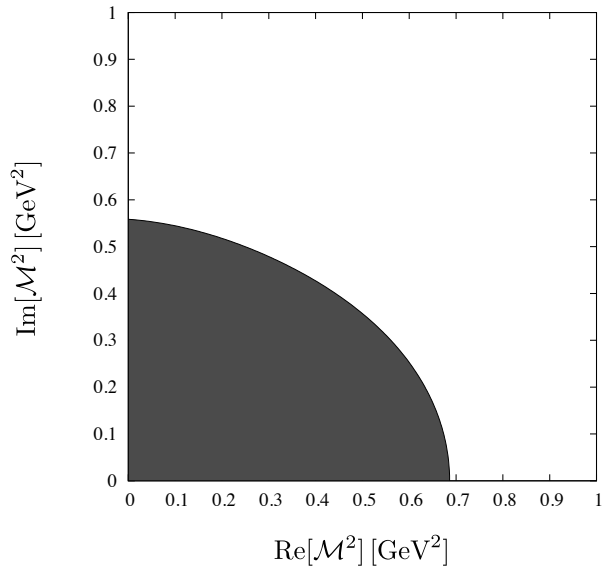}
\end{center}
 \caption{The allowed (prohibited) regions of the complex OPE 
are shown as white (black) areas in the complex $\mathcal{M}^2$ plane. 
The convergence criterion, determining the two areas and their boundary, 
is given in Eq.(\ref{eq:criterion}). }
\label{fig:inner}
\end{figure*}

We choose $r_c=0.1$ in the condition (\ref{eq:criterion}) for the domain of valid complex Borel mass.
It is then restricted outside the region specified in Fig. \ref{fig:inner}. 
\begin{table}[h]
\begin{center}
\begin{tabular*}{10cm}{@{\extracolsep{\fill}}| c||  c  c|}
\hline
$\langle    \bar q q   \rangle $                                 &  $-(0.2723 \pm  0.0018)^3 \ \mathrm{GeV}^3$  \cite{Borsanyi:2012zv}   &        \\
\hline
 $\ \ \   \langle\frac{ \alpha_s}{\pi}G ^2\rangle  \ \ \ $        &  $0.012 \pm 0.0036 \ \mathrm{GeV}^4$  \cite{Colangelo2000}      &     \\
 \hline
 $    \langle \bar{s } s \rangle      $                         &  $   (0.8\pm 0.1)\langle    \bar q q   \rangle  $   \cite{Reinders1985}         &     \\
\hline
 $m_{s}$                                                                   &  $95\pm 5\   \mathrm{MeV}  $ \cite{Beringer:1900zz}               &       \\
 \hline
 $ \kappa$                                                             &    $2 \pm 1$                              \cite{Leinweber1997}                 &        \\
 \hline
 $    \alpha_s (\mu=1\mbox{GeV})    $                         &  $   0.505 \pm  0.0167 $   \cite{Bethke:2012jm}         &     \\
\hline
\end{tabular*}
\end{center}
\caption{Values and respective uncertainties of the condensates and other parameters used for evaluating the 
OPE of Eq.(\ref{eq:realimag}).}
 \label{table:condensate}
\end{table}

\subsection{Analysis results with a single default model and $M_r^2$ value}
In our MEM analysis, we have some freedom to choose 
the default model $m(\omega)$ and the outer circle radius $M_r^2$. 
A reasonable choice for $m(\omega)$ should reflect our prior knowledge on the spectral function, 
because it gives its most probable form when no constraints from the sum rules are available. 
A suitable choice for its form 
is thus, as has been already discussed in \cite{Gubler2010}, a function which tends to zero at low energy 
and approaches the perturbative high energy limit for large $\omega$. 
A parametrization which has these properties and smoothly interpolates between the 
low and high energy limit, is given as 
\begin{equation}
\begin{split}
m_{\mathrm{step}}(\omega)=\frac{1}{4 \pi^2}(1+\frac{\alpha _s}{\pi}) \ \frac{1}{1+\mathrm{e}^{\frac{\omega _0 -\omega}{\delta}}}.\\
\end{split}
\label{eq:def}
\end{equation}
Note that $\rho(s)$ in Eq.(\ref{eq:realimag}) is dimensionless in this case and is supposed to go to the asymptotic value, $ \frac{1}{4\pi^2} (1+\frac{\alpha_s}{\pi})$ at high energy.
For a first trial, we will use 
$\omega_0 =4\  \mathrm{GeV}$ and $\delta =0.2\   \mathrm{GeV}$ in the analysis of this 
subsection and later examine the effects of other default models. 
As for $M_r^2$, it should generally not be too large because for large $M_r^2$, the damping factor of the kernels becomes weak, which 
means that the integrals over the spectral function in Eq.(\ref{eq:realimag}) 
will have large contributions from the continuum. 
$M_r^2$ should on the other hand not be taken too small to allow a sufficiently large interval above 
the prohibited region shown in Fig. \ref{fig:inner}. 
As a parameter satisfying these conditions, we choose $M_r^2 =1\,\mathrm{GeV}^2$ 
and will later investigate the effects of different choices for this value. 
For comparison, we have 
also analyzed the RBSR of the $\phi$ meson channel with MEM, 
as it was done for the $\rho$ meson in \cite{Gubler2010}. 
The Borel mass for this analysis was taken as 
$0.69 \ \mathrm{GeV}^2\leq M^2 \leq 1\  \mathrm{GeV}^2$, which corresponds to the real axis of 
the area shown in Fig. \ref{fig:inner}. 
\begin{figure*}[t]
 \begin{center}
 \includegraphics[width=14cm]{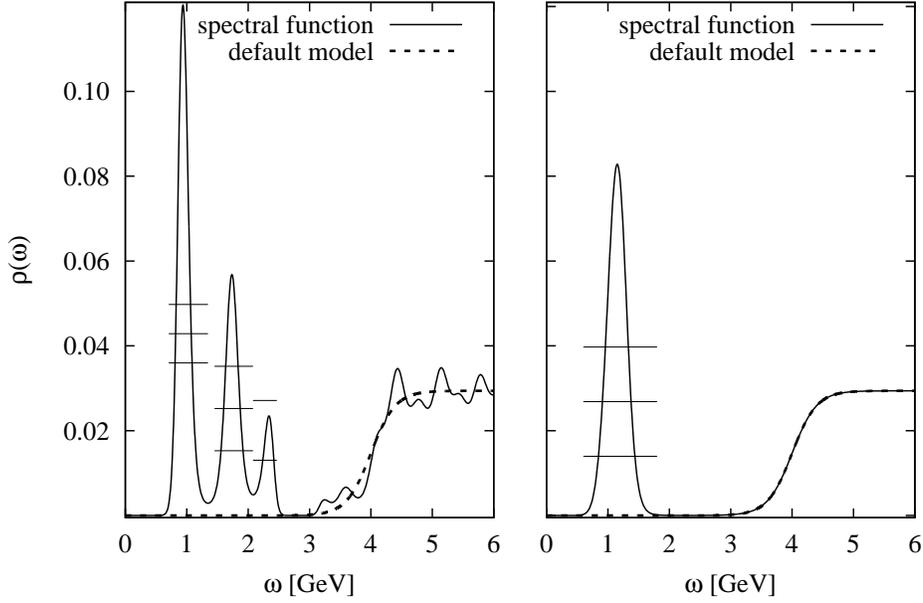}
\end{center}
 \caption{The analysis results of the CBSR (left plot) and RBSR  (right plot). 
The solid lines show the spectral function extracted from the MEM analysis. 
The dashed lines show the default model defined in Eq.(\ref{eq:def}). 
}
\label{fig:1gev}
\end{figure*}

The obtained results are shown in Fig. \ref{fig:1gev}. 
Three peaks are generated in the analysis of the CBSR. 
The estimated errors on the MEM results are shown by the three horizontal limes at each peak.
Among the three peaks, the 
first and second peak are statistically significant 
and can therefore presumably be considered to represent physical resonances. 
Further discussions on this point will follow in the next subsections. 
The third peak is on the other hand not statistically significant and hence no 
conclusions on its physical existence can be drawn. 
The positions of the first two peaks 
are given in Table\ \ref{table:p}, 
where it is observed that the peak positions agree 
quite well with the respective experimental values. 
Comparing this with the result of the 
RBSR on the right plot 
of Fig. \ref{fig:1gev}, it is seen that 
for the latter case, only one relatively broad 
peak is extracted, which can be considered to be 
a smeared version of the first two peaks obtained from the CBSR. 
This is a reasonable result, as the CBSR contains 
more detailed information on the spectral function and thus allow 
for a better resolution of its MEM extraction. 
\begin{table}[h]
\begin{center}
\begin{tabular*}{12cm}{@{\extracolsep{\fill}}c ccc}
\hline
&CBSR& RBSR &Experiment\\
\hline
\hline
1st peak\ [GeV]&0.94&1.15&1.02\\
2nd peak\ [GeV] &1.74&&1.68\\
\hline
\end{tabular*}
 \caption{Position of the peaks, extracted from the MEM analysis results 
shown in Fig. \ref{fig:1gev}. The first two columns list 
the values obtained from the complex and real Borel sum rules, while the corresponding 
experimental values are given in the third column.}
 \label{table:p}
  \end{center}
\end{table}

\subsection{Analysis results with various choices of the default model and $M_r^2$}
\begin{figure*}[!t]
 \begin{center}
  \includegraphics[scale=1.2]{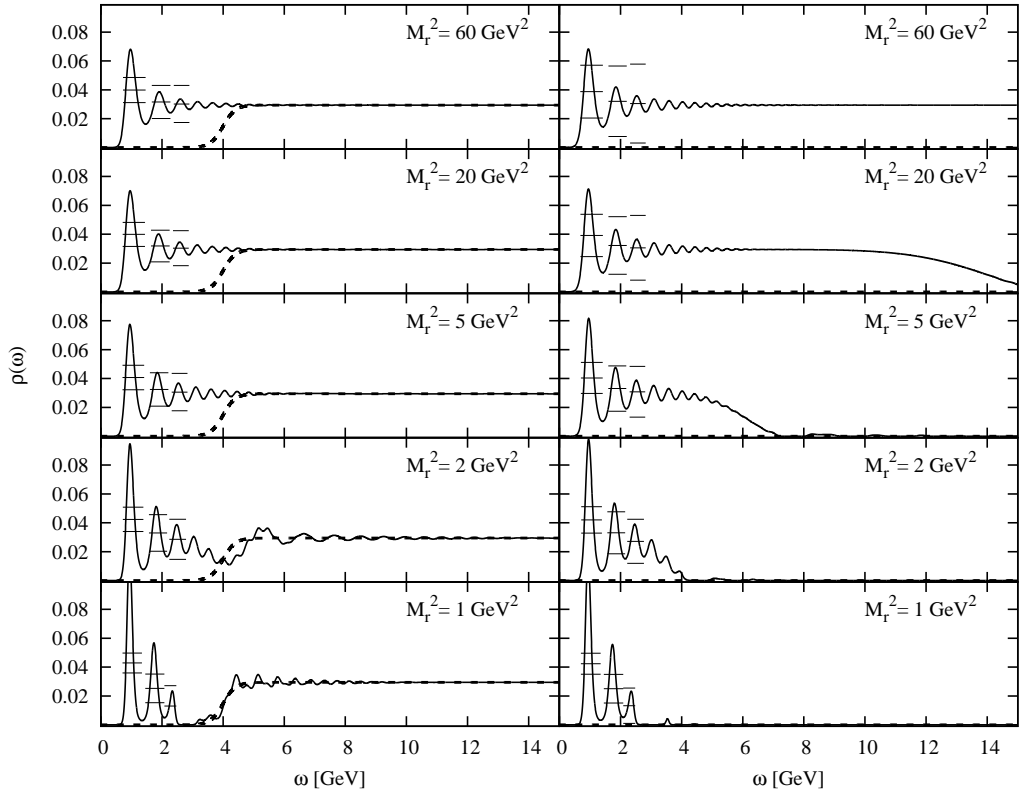}
 \end{center}
 \caption{The analysis results of the CBSR using two default models ($m_{\mathrm{step}}$ on the left side 
and $m_0$ on the right side) and five values of  $M_r^2$. 
The solid lines give the spectral function determined from the MEM analysis, while 
the dashed lines show the employed default model. }
 \label{fig:vc}
\end{figure*}

\begin{figure*}[!t]
 \begin{center}
  \includegraphics[scale=1.2]{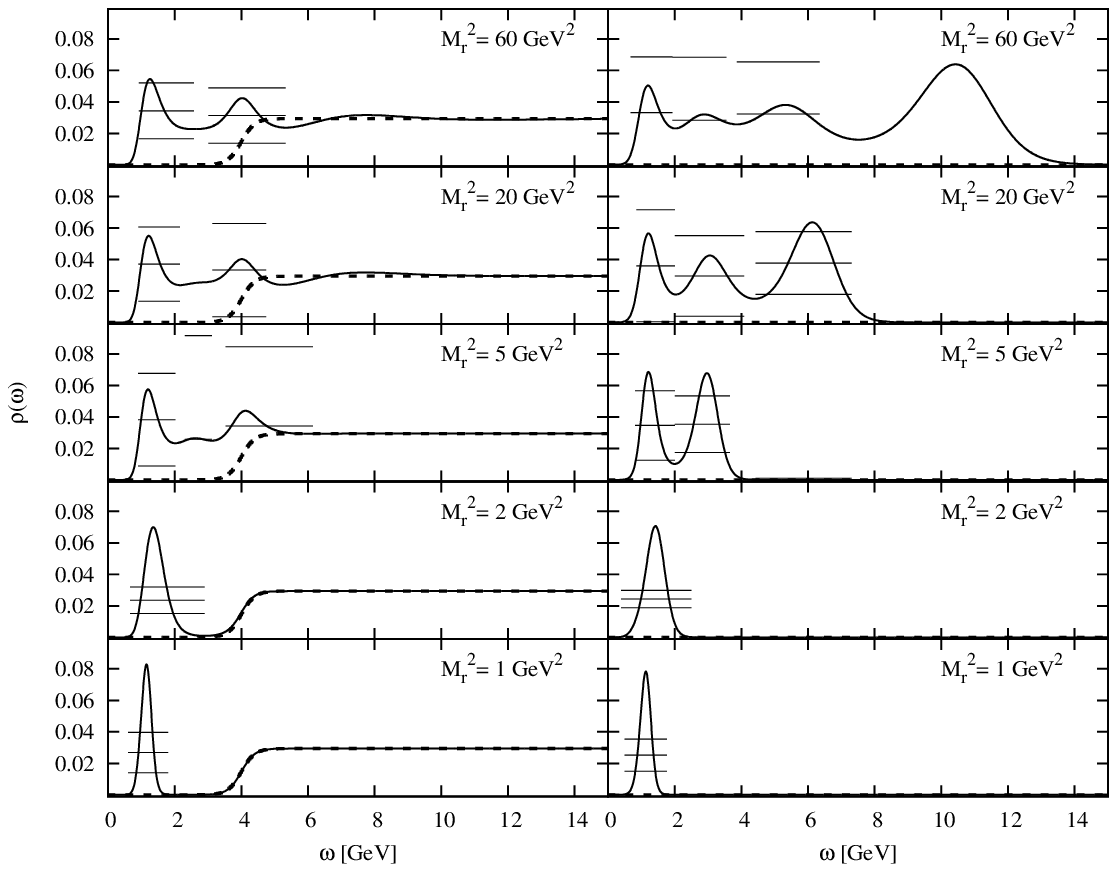}
 \end{center}
 \caption{Same as for Fig. \ref{fig:vc}, but using the RBSR as input for the MEM analysis.}
 \label{fig:vb}
\end{figure*}

To get an idea on the systematic uncertainties 
of our results, it is important to study the dependence of the 
generated spectral functions on the default model and the used value of $M_r^2$. 
We will for this purpose not only use default models of the form given in 
Eq.(\ref{eq:def}), but also another version which contains no information on 
the asymptotic behavior of the spectrum at high energy. 
The most simple form of such a default model would be just a constant, however, much smaller than the asymptotic value, $\frac{1}{4\pi ^2}(1+\frac{\alpha_s}{\pi }) \sim 0.03$.
For this reason, the following function is chosen as an alternative default model:
\begin{equation}
m_0(\omega)\ =\, 10^{-6}.
\label{eq:flat}
\end{equation}
For $M_r^2$, we take $1$, $2$, $5$, $20$ and $60$ (all in units of $[\mathrm{GeV^2}]$) 
to investigate the effect of a larger choice for this parameter. 
We have totally carried out 
ten different analyses for both the CBSR and RBSR,  
using two types of default models ($m_{\mathrm{step}}$ or $m_0$) and the above-mentioned five values of $M_r^2$. 
The corresponding results are shown in Fig.\ref{fig:vc} and Fig.\ref{fig:vb}, 
respectively. 

Let us firstly examine 
the high energy region of the obtained spectral functions. 
Looking at the results with $m_0(\omega)$ on the right side of Fig. \ref{fig:vc}, it is clear that
for larger $M_r^2$, 
the spectral functions approach the perturbative high energy limit at large $\omega$ values. 
This shows that the continuum can be reproduced by the CBSR 
irrespective of the chosen default model. 
In contrast, it is found that the reproduction of the continuum is much worse for the RBSR. 
As can be observed on the right side of Fig.\ref{fig:vb}, 
MEM tries to reproduce some sort of continuum, but can only generate a 
strongly oscillating function with a large artificial peak somewhat above 
$\omega = M_r$. 
Although the continuum is reproduced for the results with $m_{\mathrm{step}}(\omega)$ on the left side 
of Fig. \ref{fig:vb}, 
this behavior is simply a consequence of the 
default model used in this specific case. 

Next, we focus our attention on the low energy 
parts of the spectrum. 
For the lowest peak, which corresponds to the $\phi$ meson ground state, 
the obtained positions and strengths 
do not much depend on the choice of $m(\omega)$ or on the value of $M_r^2$, which 
means that the CBSR are sensitive to this state and that the MEM can extract 
its properties with only small systematic uncertainties. 
The situation is again less clear for the RBSR results of 
Fig. \ref{fig:vb}, as 
here, even the first peak is statistically significant only for $M_r^2=1\,\mbox{GeV}^2$. 

In the CBSR spectra of Fig. \ref{fig:vc}, it is furthermore seen that a statistically 
significant second peak is found for $M_r$ values up to $2\,\mbox{GeV}^2$. 
This peak however broadens and slightly moves its position as we go to 
larger values of $M_r$. It moreover looses its statistical significance for 
$M_r > 2\,\mbox{GeV}^2$ and becomes a part of the small oscillations that 
appear at the edge of the continuum for the largest few values of $M_r$. 
The properties of the second peak therefore to some degree depend on 
$M_r$ and it is not completely clear whether this peak is of physical origin or 
merely an artifact of the MEM analysis. 
We will further discuss this question in the 
the mock data analysis of the next subsection.

\subsection{Test analysis results by using the mock spectral function}
\begin{figure*}[!t]
 \begin{center}
  \includegraphics[width=14cm]{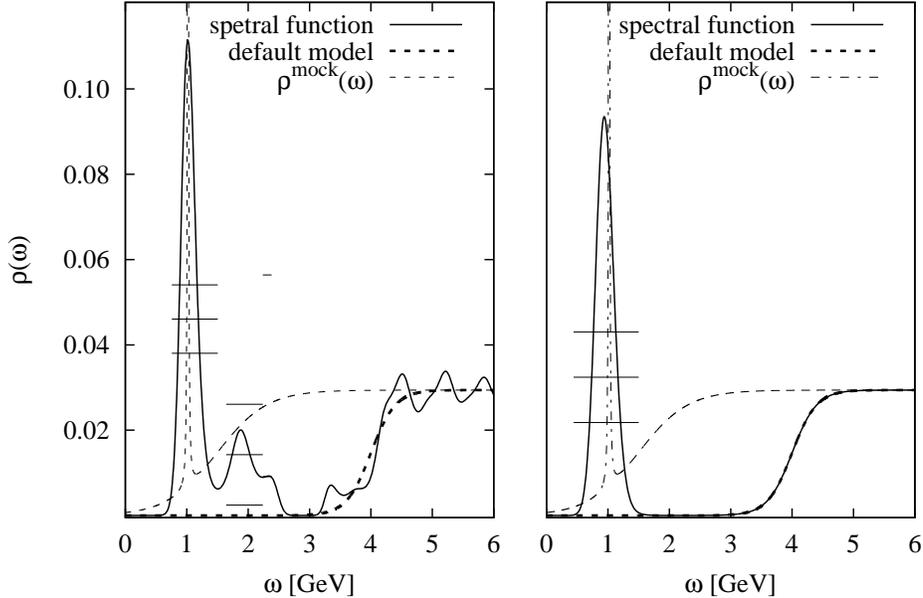}
 \end{center}
 \caption{Test analysis results of the CBSR (left plot) and RBSR  (right plot) using mock data. 
In analogy to Fig. \ref{fig:1gev}, we have employed $M_r^2=1\,\mathrm{GeV}^2$ and $m_{\mathrm{step}}$ 
for obtaining these spectra. 
The solid lines show the spectral function extracted from the MEM analysis, the thick dashed lines depict the default model
and the thin dashed lines gives the mock spectral function, $\rho^{\mathrm{mock}}$.}
 \label{fig:mock1}
\end{figure*}

\begin{figure*}[!t]
 \begin{center}
  \includegraphics[width=14cm]{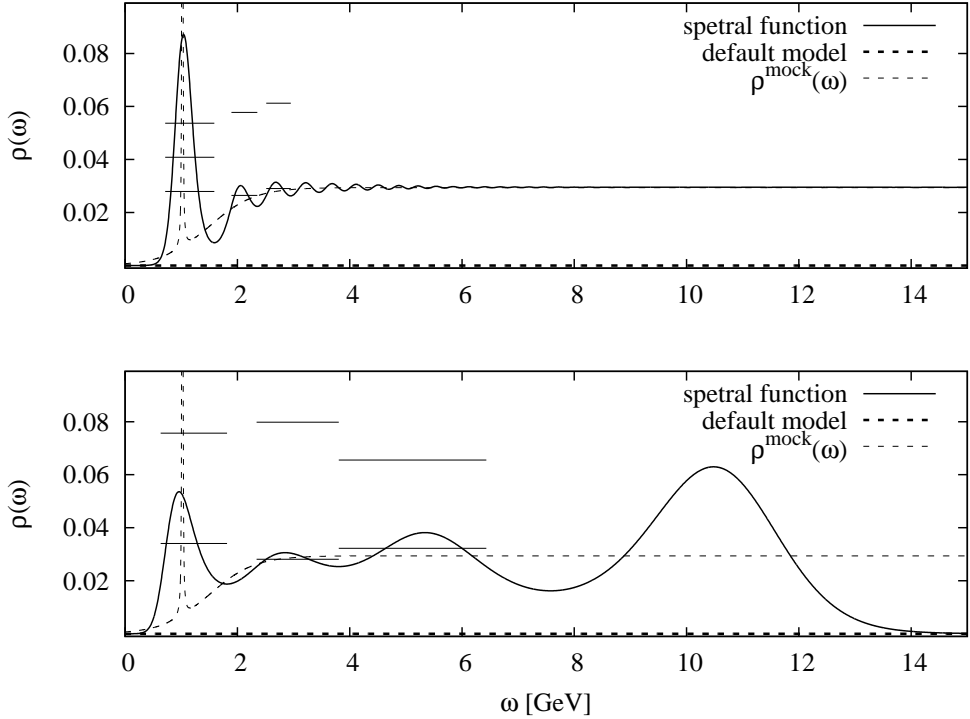}
 \end{center}
 \caption{Same as in Fig. \ref{fig:mock1}, but with $M_r^2=60\,\mathrm{GeV}^2$ and $m_{0}$. 
 The upper plot gives the result for the CBSR, while the lower plot shows the spectrum extracted from the RBSR.}
 \label{fig:mock2}
\end{figure*}

For better understanding what parts of the spectral function can be 
reliably studied with our method and what sort of artificial structures might 
appear in the MEM results, 
we have carried out a test analysis using the mock data generated from 
some specific input spectral function. 
Based on the results of this analysis, we will 
investigate whether the second peak found Figs. \ref{fig:1gev} and \ref{fig:vc} 
is physical or just an MEM artifact and will further discuss 
the different reproducibilities of the CBSR and RBSR analyses. 
For the input mock spectral function of the $\phi$-meson channel, we employ a relativistic Breit-Wigner 
peak and a smooth function describing the transition to the asymptotic value at high energies \cite{Shuryak:1993kg}, 
\begin{equation}
\rho^{\mathrm{mock}}(\omega)= \frac{3}{4\pi^2}\frac{52.4}{1+\frac{4(\omega -m)^2}{\Gamma^2} } +
\frac{1}{4 \pi^2}(1+\frac{\alpha _s}{\pi}) \ \frac{1}{1+\mathrm{e}^{\frac{\omega _0 -\omega}{\delta}}}, 
\label{eq:mockspec}
\end{equation}
which has been renormalized so that the asymptotic behavior of the spectral function at high energy is 
consistent with Eq.(\ref{eq:piOPE}), 
the OPE used in this paper. 
For the parameters appearing in Eq.(\ref{eq:mockspec}), we use the following values: 
\begin{equation}
\begin{split}
m&=1.02\ \mbox{GeV}\, ,\ \ 
\Gamma=4.26\ \mbox{MeV}\\
\omega_0&=1.5\ \mbox{GeV}\, ,\ \ 
\delta=0.4\ \mbox{GeV}\, ,\ \ 
\alpha_s=0.505
\end{split}
\end{equation}
The mock data are then numerically generated as shown below for the CBSR case:  
\begin{equation}
G^{\mathrm{mock}}(\mathcal{M}_{ij}^2) \equiv    \frac{1}{ \mathcal{M}_{ij}^2}\int^{\infty}_0  \mathrm{e}^{-\omega^2/ \mathcal{M}_{ij}^2}\rho^{\mathrm{mock}} (\omega) 2\omega d\omega.
\end{equation}
In the actual MEM analysis, we have treated $G^{\mathrm{mock}}(\mathcal{M}_{ij}^2)$ like OPE data, 
meaning that we use the the same errors and Borel mass ranges which were used in the OPE data 
analyses of the previous sections. 
Specifically, we have analyzed two cases. In the first case, we have taken 
$M_r^2\ = 1\, \mbox{GeV}^2$ and used $m_{\mathrm{step}}(\omega)$ of Eq.(\ref{eq:def})
for the default model. The respective results are shown in Fig. \ref{fig:mock1}, 
which should be compared to Fig. \ref{fig:1gev}, where the OPE data 
have been analyzed under exactly the same conditions. 
For the second case, we have used $M_r^2\ = 60\,\mbox{GeV}^2$ 
with the default model $m_{0}(\omega)$ of Eq.(\ref{eq:flat}), 
the result being shown in Fig. \ref{fig:mock2}. This case corresponds to the top right plots 
of Figs. \ref{fig:vc} and \ref{fig:vb} for the OPE data analyses. 

Firstly, let us investigate the mock data analysis results of the first case with $M_r^2=1\,\mbox{GeV}^2$. 
Looking at the left plot of Fig.\ref{fig:mock1}, it is found that for the CBSR an artificial second peak can be generated by 
the MEM analysis even if such a peak is not present in the original spectral function. 
This phenomenon can be thought of as a result of the MEM trying to reproduce the continuum, but not having 
enough information due to the small value of $M^2_r$. The extracted spectral function therefore eventually approaches 
the default model, leading to an artificial peak. 
In contrast to the OPE data analysis result of Fig. \ref{fig:1gev}, this peak is however not statistically significant. 
Also the strengths of the peaks are different: while the second peak obtained from the OPE data clearly rises 
above the continuum, the respective mock data peak does not. 
These findings show, that while it is possible that small artificial bumps or peaks can be produced by the MEM analysis, 
these will not be as large as the second peak seen in Fig.\ref{fig:1gev}. We hence can conclude that this peak 
is likely to reflect the properties of a physical state, the first excited state of the $\phi$ meson. 
As a further point, comparing both plots of Fig. \ref{fig:mock1}, it is observed that CBSR reproduces the position of the first peak 
with much better precision than the RBSR. Furthermore, considering the width of the lowest peak, it is seen that CBSR shows some 
improvement compared to the RBSR, but is nevertheless not able to reproduce the very narrow physical width of the $\phi$ meson. 

Next, we examine the results for $M_r^2=60\, \mbox{GeV}^2$. 
From Fig. \ref{fig:mock2}, we can confirm that at large energy CBSR is able to reproduce the continuum 
without relying on the default model. This is not the case for the RBSR, which instead of 
a constant behavior produces large oscillations. 
On the other hand, we found that at the lower edge of the continuum, the CBSR 
generates artificial oscillations, which  are damped out toward higher energies. 
It is hence understood that
the periodic peaks found in the OPE data analysis above the ground state peak 
for large values of $M^2_r$ (see Fig. \ref{fig:vc}) are mainly MEM artifacts. 
Even though the spectral functions obtained from the OPE data 
show a somewhat stronger second peak, which can probably be explained by the 
existence of a physical state in that region, this difference is too small to 
allow any definitive conclusions. 
To recapitulate, for large values of $M^2_r$, the sum rules 
are dominated by the continuum (which is thus well reproduced) and contains 
relatively less information on the low energy part of the spectrum. 
For obtaining information on the possible existence of excited states, one therefore 
needs to choose small enough $M^2_r$ values.

\section{Summary and conclusion}
\label{s:sum}  
We have in this study constructed the 
complex Borel plane QCD sum rules (CBSR) 
and have applied it to the $\phi$ meson channel as a first test of its validity and usefulness. 
We have explicitly demonstrated that the CBSR can be obtained by simply replacing the 
Borel mass $M^2$ of the real Borel QCD sum rules (RBSR) with a complex parameter $\mathcal{M}^2$. 
Since the Borel mass can thus take values on the 
complex plane and not only on the real axis, the CBSR allows us to extract more 
information on the spectral function than it was previously possible. 
To check whether the CBSR 
works in practice and to examine the quality of the information 
provided by the sum rules, we 
have studied the $\phi$ meson channel with the help of MEM. 
The main results of this investigation are as follows:
\begin{itemize}
\item The MEM analysis generates a clear and statistically significant lowest peak, 
which corresponds to the physical $\phi(1020)$ state. 
Its position is found to be only weakly dependent on the inputs of the MEM analysis, such as the 
default model or $M_r^2$. 
\item A second peak, which may correspond to the first excited state of the $\phi$ meson channel, 
also appears in the obtained spectral function. 
This peak is, however, only statistically significant when 
OPE data close to the origin of the complex Borel mass plane are analyzed 
(i.e. when a rather low $M_r^2$ value is used). 
Making $M_r^2$ larger, it is found to mix with artificially generated peaks 
which have no physical significance and change its position to larger energy region.
Hence, we presently can not make conclusive statements on the mass of the excited state
since it depends on the value of $M^2_r$.
Nevertheless, the fact that the peak is 
statistically significant at least for low $M_r^2$  and the results of the examination using a mock spectral function is indicative of the existence of the excited state
and warrants further studies once the OPE is known with better precision. 
\item  Besides the lowest two peaks, we have shown that the CBSR is capable 
of reproducing the continuum at high energy. Specifically, as long as $M_r^2$ is 
chosen to be sufficiently large, the MEM analysis generates the correct high energy 
 limit of the spectral function even if the default model has a different 
 limiting value. 
\end{itemize}
The CBSR hence appears to be a useful tool 
for analyzing hadronic spectral functions, which is 
superior to the conventional RBSR. 
It is important to note here that only by using MEM, we can exploit the 
full power of the CBSR, as MEM in principle allows the spectral function to 
have any specific (positive definite) form. 

\section*{Acknowledgment}
This work was partially supported by KAKENHI under Contract Nos.24540294 and 25247036.
We would like to express our sincere gratitude to Dr. Kiyoshi Sasaki for helping us with the numerical calculations.
K.O. gratefully acknowledges the support by the Japan Society for the Promotion of Science for Young Scientists (Contract No.25.6520).
\vfill\pagebreak

\appendix

\section{Derivation of dispersion relation}
\label{app:1}
\begin{figure}[!tbp]
 \begin{center}
  \includegraphics[scale=0.35]{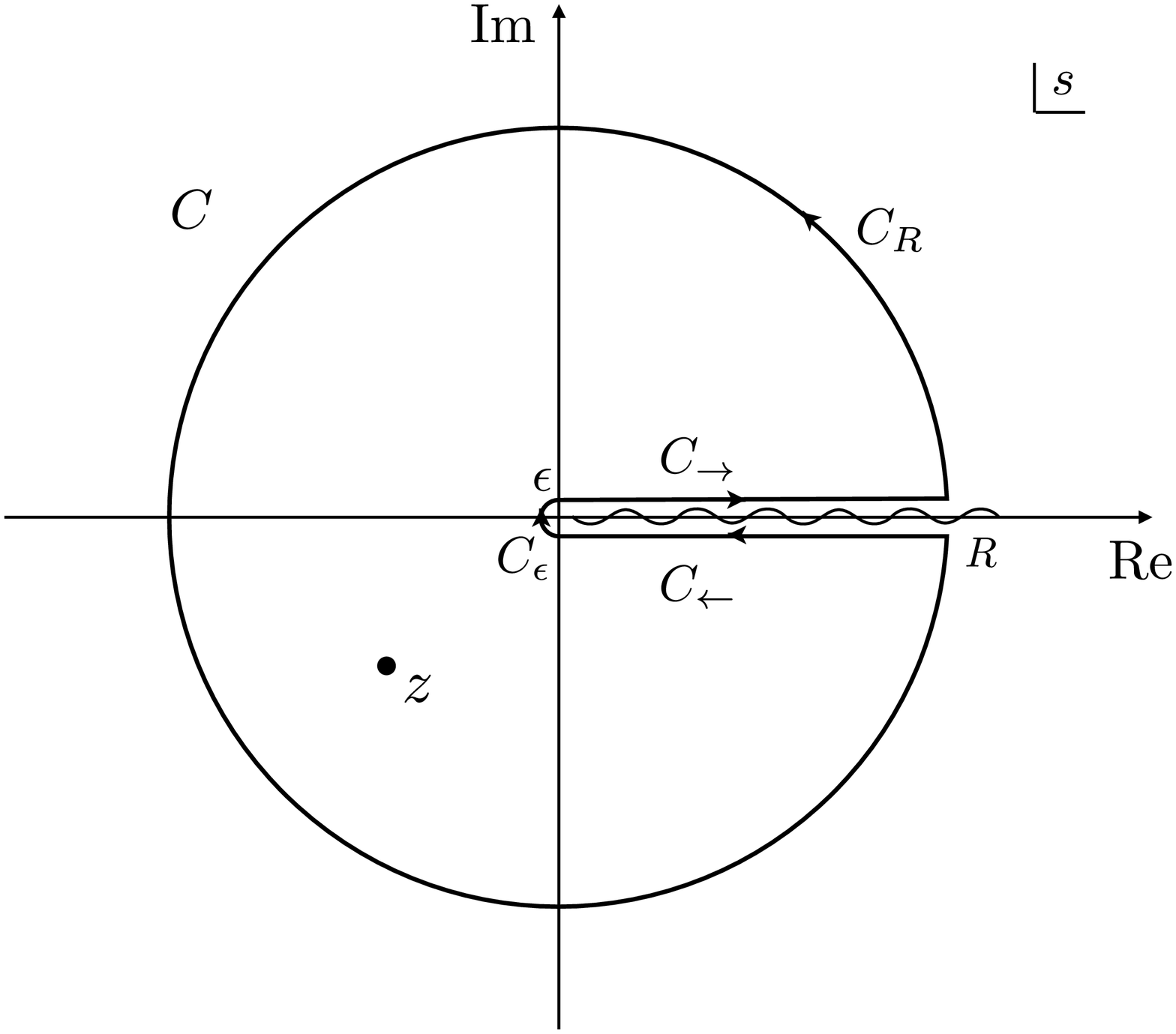}
 \end{center}
 \caption{The contour integral $C$ on the complex plane of the variable $s$, used 
in Eq.(\ref{eq:tekito-1}), divided into its segments $C_R$, $C_{\rightarrow}$, $C_{\leftarrow}$ and $C_{\epsilon}$. }
 \label{fig:derivedis}
\end{figure}

To confirm that the QCD sum rules can be formulated by complex parameters, we in this appendix 
derive the dispersion relation in detail. 
Using the analyticity of the correlator on the $q^2$ imaginary plane (excluding of course the region of positive real $q^2$), 
we can apply Cauchy's residue theorem and obtain the following equation: 
\begin{equation}
\Pi(z) = \frac{1}{2 \pi i} \int_{C}\frac{z^n \Pi(s)}{s^n(s-z)}ds. 
\label{eq:tekito-1}
\end{equation}
The contour $C$ 
is given in  Fig.\ref{fig:derivedis} and z refers to complex $q^2$. 
Firstly, the contour C is divided into $C_R + C_{\leftarrow} +C_{\epsilon} +C_{\to}$ 
as shown in Fig.\ref{fig:derivedis} and $n$ is supposed to be sufficiently large 
(but finite) so that  the integral along $C_R$ converges to zero when taking the limit 
$R \to \infty$ and $\epsilon \to 0$: 
\begin{equation}
\frac{1}{2 \pi i} \int_{C_R}\frac{z^n \Pi(s)}{s^n(s-z)}ds \xrightarrow{\epsilon \rightarrow +0,\ R\rightarrow \infty} 0.
\end{equation}
Next, the contributions of the other contour sections are considered. 
By using the equation:
\begin{equation}
\frac{z^n}{s^n (s-z)} =\frac{1}{s-z} - \sum_{k=0} ^{n-1} \frac{z^k}{s^{k+1} },
\label{eq:formula}
\end{equation}
the integrals along $C_{\rightarrow }$ and $C_\leftarrow$ can be divided into terms containing only polynomials in z and 
terms with other analytical structures:
\begin{equation}
\begin{split}
&\frac{1}{2 \pi i}  \int_{C_\rightarrow}\frac{z^n \Pi(s)}{s^n(s-z)}ds  \ \   +   \ \   \frac{1}{2 \pi i} \int_{C_\leftarrow}\frac{z^n \Pi(s)}{s^n(s-z)}ds\\
=\,&  \frac{1}{2 \pi i}  \int_{C_\rightarrow}\frac{ \Pi(s)}{s-z}ds  \ \   +   \ \   \frac{1}{2 \pi i} \int_{C_\leftarrow}\frac{\Pi(s)}{s-z}ds \ \ +\ \ \mbox{polynomial \ in\ z}        \\
=\,&  \frac{1}{2 \pi i} \int_{0}^{R}     \frac{  \Pi(s+ i \epsilon)  }{s+i\epsilon-z }  ds   \ \ +\ \  \frac{1}{2 \pi i} \int_{R}^{0}     \frac{  \Pi(s- i \epsilon)  }{s-i\epsilon-z }  ds   \ \ +\ \ \mbox{polynomial in z} \\ 
=\,&  \frac{1}{2 \pi i} \int_{0}^{R}     \frac{     (s-i\epsilon -z)      \Pi(s+ i \epsilon)       -(s+i\epsilon -z)      \Pi(s-i \epsilon)          }{(s-z)^2 }  ds     \ \ +\ \ \mbox{polynomial in z} \\ 
=\,&  \frac{1}{\pi } \int_{0}^{R}     \frac{       \mbox{Im}\Pi(s+ i \epsilon)          }{s-z }  ds \ \ -\ \   \frac{\epsilon}{\pi } \int_{0}^{R}     \frac{       \mbox{Re}\Pi(s+ i \epsilon)          }{(s-z)^2 }  ds     \ \ +\ \ \mbox{polynomial in z} \\ 
&\xrightarrow{\epsilon \rightarrow +0,\ R\rightarrow \infty} \int_{0}^{\infty} \frac{\rho(s)}{s-z}ds  \ \ +\ \   \mathrm{polynomial\  in\  z}
\end{split}
\end{equation}
For treating 
the integral along the small half circle $C_{\epsilon}$, we again use Eq.(\ref{eq:formula}) to obtain 
\begin{equation}
\begin{split}
\frac{1}{2 \pi i}\int_{C_{\epsilon}}\frac{z^n \Pi(s)}{s^n(s-z)}ds & =  \frac{\epsilon}{2 \pi}  \int_{\frac{3 \pi}{2}} ^{\frac{\pi}{2} }  \frac{ \Pi(  \epsilon \mathrm{e}^{i \theta}   )}{\epsilon \mathrm{e}^{i \theta} -z} \mathrm{e}^{i \theta} d\theta
    \ \ +\ \  \mathrm{polynomial\ in\ z} \\
    & \xrightarrow{\epsilon \rightarrow +0} \mathrm{polynomial\ in\ z}.
\end{split}
\end{equation}
Note that we have assumed here  that Го(s) does not have a singularity at $s=0$. 
Adding the various contributions, we finally arrive at the desired dispersion relation: 
\begin{equation}
\Pi(z) =\int_{0}^{\infty} \frac{\rho(s)}{s-z}ds  \ \ +\ \   \mathrm{polynomial\  in\  z}.
\label{eq:kekka}
\end{equation}
The important point here is that the above equation has been derived without restricting  z to real numbers. 
It also should be mentioned that although we have chosen the 
contour $C$ not to include 
the origin, we as well could have taken the origin to lie inside the contour. 
In that case, $- \sum_{k=0}^{n-1}\frac{\Pi^{(n)}(0)}{k!}z^k$ would have to be 
added to Eq.(\ref{eq:tekito-1}). The final result, however, does not change its form since 
this additional term is just another 
polynomial in $z$  and is thus consistent with Eq.(\ref{eq:kekka}). 

\section{Proof of the interchangeability of the Borel transformation interchange with the integral over $s$} 
In this appendix, we will prove that for $\mathrm{Re}[\mathcal{M}^2]>0$  it is possible to interchange the Borel transformation with the integral 
over $s$ in Eq. (\ref{eq:keisan}). 
For doing this, we rely on the following theorem \cite{kodaira2003}: 

\begin{theorem}\label{T0.1}
The sequence of continuous functions $\{ f_n(x)\}$ is defined in $I=(a,\infty)$. 
It is assumed that the limit $f(x) = \lim_{n \to \infty} f_n(x)$ exists and that it 
is continuous in $I$. 
Furthermore, a function $g(x)$ which satisfies the following conditions is assumed to exist:
\begin{itemize}
\item for all $n \in \mathbb{N} $ and $x\in I$  \  $|f_n(x)| \leq g(x)$
\item $\displaystyle{\int_a^{\infty} g(x)<\infty}$
\end{itemize}
When these conditions hold, 
the limit of taking $n$ to infinity 
can be interchanged with the integral over $x$ as shown below. 
\begin{equation}
\lim_{n\to \infty} \int_{a}^{\infty} f_n(x)dx = \int_{a}^{\infty}\Bigl(  \lim_{n\to\infty} f_n(x) \Bigr)dx
=\int_{a}^{\infty} f(x)dx
\end{equation}
\end{theorem}

In order to apply this theorem to the problem at hand, we use the last line of 
Eq.(\ref{eq:keisan}) and rewrite it as follows: 
\begin{equation}
\hat{B}_{[|z|]} \int_{0}^{\infty}\frac{\rho(s)}{s-z}ds  =\lim_{n\rightarrow \infty}\int_{0}^{\infty}F_n(s) ds,
\end{equation}
with 
\begin{equation}
F_n(s)=\frac{1}{M^2 \mathrm{e}^{i \theta}} \Bigl(1+ \frac{1}{n}\frac{s}{M^2 \mathrm{e}^{i \theta}  } \Bigr)^{-(n+1)}\rho(s).
\end{equation}
To be able 
to interchange the Borel transformation with the integral therefore would mean that 
\begin{equation}
\begin{split}
\lim_{n\rightarrow \infty}\int_{0}^{\infty}\mathrm{Re}[F_{n}(s)]ds= \int_{0}^{\infty} \lim_{n\rightarrow \infty}{\mathrm{Re}}[F_{n}(s)]ds,   \\
\lim_{n\rightarrow \infty}\int_{0}^{\infty}\mathrm{Im}[F_{n}(s)]ds= \int_{0}^{\infty} \lim_{n\rightarrow \infty}{\mathrm{Im}}[F_{n}(s)]ds,
\end{split}
\label{eq:koukan}
\end{equation}
hold. Both relations can be proven for $\cos{\theta}>0$, 
for which one has to show that the assumptions needed for the above theorem are satisfied. 
For definiteness, we here only derive the first relation and note that the 
second one can be treated analogously. 
Firstly, it is clear that the limit $\lim_{n \to \infty} \mathrm{Re}[F_{n}(s)]$ 
exists and is continuous. This limit is given in Eq.(\ref{eq:reker}). 
Next, the absolute value of the real part of $F_n(s)$ can be estimated as follows: 
\begin{equation}
\begin{split}
|\mathrm{Re}[F_n(s)]  |&  \leq |F_n(s)|\\
& = \frac{1}{M^2}\Bigl( 1+\frac{2s\cos \theta  }{nM^2}  +     \frac{s^2}{n^2 M^4 } \Bigr  )^{-\frac{n+1}{2}}\rho(s)     \\
&\leq \frac{1}{M^2}\Bigl( 1+\frac{2s\cos \theta  }{nM^2}   \Bigr )^{-\frac{n}{2}}\rho(s)
\end{split}
\end{equation}
As $(1+a/x)^{-x}$ for $a>0$ is a monotonously decreasing function of $x$, we obtain 
\begin{equation}
|\mathrm{Re}[F_n(s)]  |   \leq  \frac{1}{M^2}\Bigl( 1+\frac{2s\cos \theta  }{N M^2}   \Bigr )^{-N/2}\rho(s)  \ \ \ \ \ \  (\mathrm{for \ all \ n \geq N}),
\label{eq:condition1}
\end{equation}
where $N$ is a natural number. 
It should be kept in mind here that this inequality is only valid for $\cos\theta>0$. 
We can take $N$ to be finite so that
\begin{equation}
  \int_0^{\infty}  \frac{1}{M^2}         \frac{\rho(s)}{   ( 1+\frac{2s\cos \theta  }{N M^2}   )^{N/2}  }   ds \ <\infty, 
\label{condition2}
\end{equation}
because the behavior of $\rho(s)$ in the high energy region is supposed to be polynomial.
Redefining $\mathrm{Re}[F_n(s)]  $ by taking $n=n+N$ (which does not matter since we are intereted in the limit $n\to \infty$), 
we can write 
\begin{equation}
|\mathrm{Re}[F_n(s)] | \leq  \frac{1}{M^2}         \frac{\rho(s)}{   ( 1+\frac{2s\cos \theta  }{N M^2}   )^{N/2}  }   \ \ \ \ \ \  (\mathrm{for \ all \ n} ). 
\end{equation}
Identifying  $\frac{\rho(s)}{   ( 1+\frac{2s\cos \theta  }{N M^2}   )^{N/2}  }$ with the function $g$ of the theorem, the proof is complete. 
As mentioned above, 
the proof of the second equation in Eq.(\ref{eq:koukan}) can be done in the same way. 
We have hence proven that one can interchange the Borel transformation with the integral over $s$ for $\cos{\theta}>0$. 

\section{Borel transformation on the complex plane}
In this appendix, we derive the Borel transformations of the complex functions given in section \ref{effectivity}, i.e. 
\begin{eqnarray}
\label{eq:f1p}
\hat{B}_{[|z|]}\,z^k&=&0,\\
\label{eq:f2p}
\ \hat{B}_{[|z|]} \,\Bigl( \frac{1}{z} \Bigr)^k &=&\frac{(-1)^k}{(k-1)!}\Bigl(\frac{1}{M^2\mathrm{e}^{i\theta}} \Bigr)^k,\\
\label{eq:f3p}
\hat{B}_{[|z|]} \,z^k \ln{\Bigl(-\frac{z}{{\mu}^2} \Bigr)} &=& -k!(M^2\mathrm{e}^{i\theta})^k,\\
\label{eq:f4p}
\hat{B}_{[|z|]} \,\Bigl(\frac{1}{s-z}\Bigr)^k &=&\frac{1}{(k-1)!}\frac{1}{(M^2\mathrm{e}^{i\theta})^k}  \mathrm{e}^{-s/( M^2 \mathrm{e}^{i\theta})}. 
\end{eqnarray}
Although it is possible to derive them following the definition of $\hat{B}_{[|z|]}$, we for simplicitly utilize the following formulae 
for the Borel transformation of real functions ($\theta=0$). 
\begin{eqnarray}
\label{eq:f5p}
\hat{B}_{[|z|]}\,|z|^k&=&0,\\
\label{eq:f6p}
\hat{B}_{[|z|]} \,\Bigl( \frac{1}{|z|} \Bigr)^k &=&\frac{1}{(k-1)!}\Bigl(\frac{1}{M^2} \Bigr) ^k,\\
\label{eq:f7p}
\hat{B}_{[|z|]} \,|z|^k \ln{ \Bigl( \frac{|z|}{{\mu}^2} \Bigr)} &=& -(-1)^kk!(M^2)^k,\\
\label{eq:f8p}
\hat{B}_{[|z|]}\, \Bigl( \frac{1}{s+|z|} \Bigr)^k&=&\frac{1}{(k-1)!}\frac{1}{(M^2)^k}\mathrm{e}^{-s/M^2}, 
\end{eqnarray}
which are consistent with Eq.(\ref{eq:fr}), once $|z|$ is replaced with $-q^2$. 
The strategy is to separate the phase of $z$ by substituting $z=|z|\mathrm{e}^{i(\theta-\pi)}$, after which 
the calculations are straightforward. The first relations is easily derived, 
\begin{equation}
\begin{split}
\hat{B}z^k=(-\mathrm{e}^{ik\theta})^k\hat{B}_{[|z|]}|z|^k=0, 
\end{split}
\end{equation}
while the second one is obtained as
\begin{equation}
\begin{split}
\hat{B}_{[|z|]} \Bigl( \frac{1}{z} \Bigr)^k&=\frac{1}{(-\mathrm{e}^{i\theta})^k}\hat{B}_{[|z|]}\frac{1}{|z|^k}\\
&=\frac{1}{(-\mathrm{e}^{i\theta})^k}  \frac{1}{(k-1)!}  \frac{1}{(M^2)^k}\\
&= \frac{(-1)^k}{(k-1)!}  \frac{1}{(M^2\mathrm{e}^{i\theta})^k} .
\end{split}
\end{equation}
The third and fourth equations are derived in a similar way: 
\begin{equation}
\begin{split}
\hat{B}_{[|z|]} z^k \ln{   \Bigl(- \frac{z}{\mu ^2}   \Bigr )}  &=  \hat{B}_{[|z|]}  (-\mathrm{e}^{i\theta})^k    |z|^k    \ln{   \Bigl(\frac{|z| \mathrm{e}^{i\theta}}{\mu ^2}    \Bigr)}                \\
&=   (-\mathrm{e}^{i\theta})^k \Bigl[  \ \    \hat{B}_{[|z|]}    |z|^k    \ln{   \Bigl(\frac{|z| }{\mu ^2}    \Bigr)}+i\theta \hat{B}_{[|z|]}  |z|^k      \ \    \Bigr] \\
&= (-\mathrm{e}^{i\theta})^k    [-(-1)^k k! (M^2)^k]\\
&=-k! (M^2\mathrm{e}^{i\theta} )^k,
\end{split}
\end{equation}

\begin{equation}
\begin{split}
\hat{B}_{[|z|]} \,\Bigl(\frac{1}{s-z}\Bigr)^k &=(\mathrm{e}^{-i\theta})^k\hat{B}_{[|z|]}\, \Bigl(\frac{1}{(s\mathrm{e}^{-i\theta})+|z|}\Bigr)^k\\
&=(\mathrm{e}^{-i\theta})^k \frac{1}{(k-1)!} \frac{1}{(M^2)^k}\mathrm{e}^{-(s\mathrm{e}^{-i\theta})/M^2}\\
&=\frac{1}{(k-1)!}\frac{1}{(M^2\mathrm{e}^{i\theta})^k}  \mathrm{e}^{-s/( M^2 \mathrm{e}^{i\theta})}.
\end{split}
\end{equation}

\clearpage
\bibliographystyle{h-physrev3}
\bibliography{reference.Gubler}

\end{document}